\documentclass[conference]{IEEEtran}
\IEEEoverridecommandlockouts
\usepackage{cite}
\usepackage{amsmath,amssymb,amsfonts}
\usepackage{algorithmic}
\usepackage{graphicx}
\usepackage{textcomp}
\usepackage{xcolor}
\usepackage{siunitx}
\usepackage{amssymb}
\usepackage{bbm} 
\usepackage[nocomma]{optidef}
\usepackage{tikz}
\usepackage{pgfplots}
\usepackage{hyperref}
\usepackage[font=scriptsize,skip=0pt]{caption}
\usepackage{subcaption}
\usepackage{float}
\usepackage{booktabs}
\usepackage[linesnumbered,ruled,vlined]{algorithm2e}
\SetKwInput{KwInput}{Input}
\SetKwInput{KwOutput}{Output}
\def\BibTeX{{\rm B\kern-.05em{\sc i\kern-.025em b}\kern-.08em
    T\kern-.1667em\lower.7ex\hbox{E}\kern-.125emX}}

    \usepackage{acronym}
    \acrodef{ac}[AC]{Admission Control}
    \acrodef{acm}[ACM]{Admission Control Module}
    \acrodef{acr}[AR]{Acceptance Ratio}
    \acrodef{aco}[ACO]{Ant Colony Optimization}
    \acrodef{adc}[AC]{Admission Control}
    \acrodef{af}[AF]{Application Function}
    \acrodef{api}[API]{Application Programming Interface}
    \acrodef{aoi}[AoI]{Age of Information}
    \acrodef{ar}[AR]{Augmented Reality}
    \acrodef{awgn}[AWGN]{Additive White Gaussian Noise}
    \acrodef{be}[BE]{Best Effort}
    \acrodef{bs}[BS]{Base Station}
    \acrodef{bbu}[BBU]{Baseband Unit}
    \acrodef{bnn}[BNN]{Bayesian Neural Network}
    \acrodef{bpsk}[BPSK]{Binary Phase Shift Keying}
    \acrodef{camra}[CAMRA]{Context-Aware Multi Resource Allocation}
    \acrodef{cc}[CC]{Cloud Computing}
    \acrodef{ceu}[CEU]{Central Edge Unit}
    \acrodef{cmab}[C-MAB]{Contextual Multi-Armed Bandit}
    \acrodef{cn}[CN]{Core Network}
    \acrodef{cnf}[CNF]{Cloud Native Function}
    \acrodef{cns}[CNS]{Cloud Network Slicing}
    \acrodef{cpu}[CPU]{Central Processing Unit}
    \acrodef{csp}[CSP]{Communication Service Provider}
    \acrodef{cu}[CU]{Central Unit}
    \acrodef{dc}[DC]{Data Center}
    \acrodef{dp}[DP]{Dynamic Programming}
    \acrodef{dl}[DL]{Deep Learning}
    \acrodef{drl}[DRL]{Deep Reinforcement Learning}
    \acrodef{dnn}[DNN]{Deep Neural Network}
    \acrodef{du}[DU]{Distributed Unit}
    \acrodef{e2e}[E2E]{End-to-End}
    \acrodef{ec}[EC]{Edge Computing}
    \acrodef{embb}[eMBB]{enhanced Mobile Broadband}
    \acrodef{en}[EN]{Edge Node}
    \acrodef{ens}[ENS]{Edge Network Slicing}
    \acrodef{epc}[EPC]{Evolved Packet Core}
    \acrodef{es}[ES]{Edge Server}
    \acrodef{etsi}[ETSI]{European Telecommunications Standards Institute}
    \acrodef{exprp}[ExpRP]{Exponential Reservation Policy}
    \acrodef{fcfs}[FCFS]{First Come First Serve}
    \acrodef{fh}[FH]{Fronthaul}
    \acrodef{fn}[FN]{Fog Node}
    \acrodef{fomdmkp}[FOMDMKP]{Fractional Online  Multi-Dimensional Multiple Knapsack Problem}
    \acrodef{fpga}[FPGA]{Field-Programmable Gate Array}
    \acrodef{gpp}[GPP]{General Purpose Processor}
    \acrodef{gppj}[GPP]{Generation Partnership}
    \acrodef{gpu}[GPU]{Graphic Processor Unit}
    \acrodef{gsm}[GSM]{Global System for Mobile Communications}
    \acrodef{ip}[IP]{Integer Program}
    \acrodef{ilp}[ILP]{Integer Linear Program}
    \acrodef{inp}[InP]{Infrastructure Provider}
    \acrodef{iot}[IoT]{Internet-of-Things}
    \acrodef{itu}[ITU]{International Telecommunications Union}
    \acrodef{kpi}[KPI]{Key Performance Indicators}
    \acrodef{lp}[LP]{Linear Program}
    \acrodef{linrp}[LinRP]{Linear Reservation Policy}
    \acrodef{lmgtfy}[LMGTFY]{Let Me Google That For You}
    \acrodef{lxc}[LXC]{Linux Container}
    \acrodef{mab}[MAB]{Multi-Armed Bandit}
    \acrodef{mac}[MAC]{Medium Access Protocol}
    \acrodef{mano}[MANO]{Management \& Orchestration}
    \acrodef{mcs}[MCS]{Modulation and Coding Scheme}
    \acrodef{mdp}[MDP]{Markov Decision Process}
    \acrodef{mdkp}[MdKP]{Multidimensional Knapsack Problem}
    \acrodef{mec}[MEC]{Mobile Edge Computing}
    \acrodef{mimo}[MIMO]{Multiple Input Multiple Output}
    \acrodef{milp}[MILP]{Multiple Integer Linear Program}
    \acrodef{ml}[ML]{Machine Learning}
    \acrodef{mm}[MM]{Markov Model}
    \acrodef{mmtc}[mMTC]{massive Machine Type Communication}
    \acrodef{mno}[MNO]{Mobile Network Operator}
    \acrodef{naas}[NaaS]{Network-as-a-Service}
    \acrodef{nf}[NF]{Network Function}
    \acrodef{nfv}[NFV]{Network Function Virtualization}
    \acrodef{ngmn}[NGMN]{Next-Generation Mobile Networks}
    \acrodef{nr}[NR]{New Radio}
    \acrodef{nrb}[NRB]{Network Resource Broker}
    \acrodef{ns}[NS]{Network Slicing}
    \acrodef{nsl}[NSL]{Network Slice}
    \acrodef{nsb}[NSB]{Network Slice Broker}
    \acrodef{nse}[NSE]{Network Service Embedding}
    \acrodef{nsp}[NSP]{Network Slice Provider}
    \acrodef{nst}[NST]{Network Slice Template}
    \acrodef{nsr}[NSR]{Network Slice Request}
    \acrodef{nsrp}[NSRP]{Network Slice Request Placement}
    \acrodef{osacp}[OSACP]{Online Slice Admission Control Problem}
    \acrodef{osp}[OSP]{Online Slice Placement}
    \acrodef{oens}[OENS]{Online Edge Node Selection}
    \acrodef{ofns}[OFNS]{Online Fog Node Selection}
    \acrodef{ofdma}[OFDMA]{Orthogonal Frequency-Division Multiple Access}
    \acrodef{omdkp}[OMdKP]{Online Multidimensional Knapsack Problem}
    \acrodef{omdmkp}[OMdMKP]{Online Multi-dimensional Multiple Knapsack Problem}
    \acrodef{okp}[OKP]{Online Knapsack Problem}
    \acrodef{osac}[OSAC]{Online Slice Admission Control}
    \acrodef{osp}[OSP]{Online Slice Placement}
    \acrodef{pdcp}[PDCP]{Packet Data Convergence Protocol}
    \acrodef{pnf}[PNF]{Physical Network Function}
    \acrodef{prb}[PRB]{Physical Resource Block}
    \acrodef{qam}[QAM]{Quadrature Amplitude Modulation}
    \acrodef{qoe}[QoE]{Quality of Experience}
    \acrodef{qos}[QoS]{Quality of Service}
    \acrodef{qpsk}[QPSK]{Quadrature Phase Shift Keying}
    \acrodef{ra}[RA]{Resource Allocation}
    \acrodef{rac}[RAC]{Radio Admission Control}
    \acrodef{ram}[RAM]{Random Access Memory}
    \acrodef{ran}[RAN]{Radio Access Network}
    \acrodef{rans}[RaNS]{Radio Network Slicing}
    \acrodef{resb}[ResB]{Resource Broker}
    \acrodef{rb}[RB]{Resource Block}
    \acrodef{rl}[RL]{Reinforcement Learning}
    \acrodef{rlc}[RLC]{Radio Link Control}
    \acrodef{rrc}[RRC]{Radio Resource Control}
    \acrodef{rrm}[RRM]{Radio Resource Management}
    \acrodef{ru}[RU]{Radio Unit}
    \acrodef{sac}[SAC]{Slice Admission Control}    
    \acrodef{sba}[SBA]{Service-Based Architecture}
    \acrodef{scfdma}[SCFDMA]{Single Carrier-Frequency-Division Multiple Access}
    \acrodef{sdc}[SDC]{Software Defined Controller}    
    \acrodef{sefc}[SEFC]{Serverless Functions Chain}
    \acrodef{sdn}[SDN]{Software Defined Networking}
    \acrodef{sdnc}[SDNC]{Software Defined Network Controller}
    \acrodef{sdr}[SDR]{Software Defined Radio}
    \acrodef{sf}[SF]{Serverless Functions}
    \acrodef{sfc}[SFC]{Service Function Chain}
    \acrodef{siso}[SISO]{Single Output Single Input}
    \acrodef{sla}[SLA]{Service Level Agreement}    
    \acrodef{soa}[SoA]{State-of-the-Art}
    \acrodef{sr}[SR]{Service Request}
    \acrodef{sla}[SLA]{Service Level Agreement}
    \acrodef{slaas}[SlaaS]{Slice as a Service}
    \acrodef{slr}[SLR]{Slice Request}
    \acrodef{so}[SO]{Slice Orchestrator}
    \acrodef{sp}[SP]{Slice Placement}
    \acrodef{st}[ST]{Slice Tenant}
    \acrodef{tns}[TNS]{Transport Network Slicing}
    \acrodef{ucb}[UCB]{Upper Confidence Bound}
    \acrodef{ue}[UE]{User Equipment}
    \acrodef{urllc}[uRLLC]{ultra-Reliable and Low-Latency Communication}
    \acrodef{v2x}[V2X]{Vehicle-to-Everything}
    \acrodef{vec}[VEC]{Virtual Edge Computing}
    \acrodef{vim}[VIM]{Virtualized Infrastructure Manager}
    \acrodef{vm}[VM]{Virtual Machine}
    \acrodef{vne}[VNE]{Virtual Network Embedding}
    \acrodef{vnf}[VNF]{Virtual Network Function} 
    \acrodef{vnr}[VNR]{Virtual Network Request}
    \acrodef{vr}[VR]{Virtual Reality}
    \acrodef{vran}[vRAN]{Virtual Radio Access Network}
    \acrodef{wtp}[WTP]{Willingness To Pay}
    \acrodef{wtpr}[WTPR]{Willingness-To-Pay Ratio}
\begin{document}

\title{An Online Multi-dimensional Knapsack Approach for Slice Admission Control}

\author{
\IEEEauthorblockN{Jesutofunmi Ajayi, 
                      Antonio Di Maio, 
                      Torsten Braun, 
                      Dimitrios Xenakis 
    }}

\maketitle

\begin{abstract}
        \acl{ns} has emerged as a powerful technique to enable cost-effective, multi-tenant communications and services over a shared physical mobile network infrastructure.
        One major challenge of service provisioning in slice-enabled networks is the uncertainty in the demand for the limited network resources that must be shared among existing slices and potentially new \aclp{nsr}.
        In this paper, we consider admission control of \aclp{nsr} in an online setting, with the goal of maximizing the long-term revenue received from admitted requests.
        We model the \acl{sac} problem as an \acl{omdkp} and present two reservation-based policies and their algorithms, which have a competitive performance for \aclp{omdkp}.
        Through Monte Carlo simulations, we evaluate the performance of our online admission control method in terms of average revenue gained by the \acl{inp}, system resource utilization, and the ratio of accepted slice requests. 
        We compare our approach with those of the online \acl{fcfs} greedy policy.
        The simulation's results prove that our proposed online policies increase revenues for \aclp{inp} by up to 12.9\% while reducing the average resource consumption by up to 1.7\%
        In particular, when the tenants' economic inequality increases, an \acl{inp} who adopts our proposed online admission policies gains higher revenues compared to an \acl{inp} who adopts \acl{fcfs}.
                
    \end{abstract}


\section{Introduction} \label{sec:intro}
    \ac{ns} has emerged as a powerful new technique that supports cost-effective, multi-tenant communication and computation over a shared physical network infrastructure.
    \ac{ns} is enabled by increasing the virtualization of next-generation mobile networks, which is provided by a combination of \ac{nfv} and \ac{sdn}.
    Through network virtualization, a higher level of network flexibility can be achieved by the elastic (re)allocation of infrastructure resources that are required by the heterogeneous applications and services served by network slices and deployed in 5G networks~\cite{9737314}. 
    
    A major challenge in the management of slice-enabled 5G networks is the resource allocation to support dynamic and online (or real-time) provisioning of network slices.
    Addressing this problem requires efficient and timely assignment of virtualized network resources to optimize network objectives such as resource utilization, the \ac{qos}, or the number of admitted slices in the network~\cite{9737314,8678397}. 
    Hence, an important consideration in the dynamic provisioning of network slices is the decision that an \ac{inp} needs to make when an admission request (i.e., \ac{nsr}) from a tenant is received for the deployment of a new network slices onto the physical mobile infrastructure~\cite{8480646}. 
    This is commonly described as the \acf{sac} problem, which requires finding an optimal strategy that achieves the \acp{inp}'s primary objectives, while also considering the Spatio-temporal traffic dynamics of network slices~\cite{bakri2021}.
    While the \ac{sac} problem is challenging to address as it typically involves a level of uncertainty in terms of the arrival rate, resource requirements, expected revenue, and the lifetime of \acp{nsr}, devising an appropriate \ac{sac} mechanism is important for achieving improved efficiency and fairness in a shared network infrastructure~\cite{9690624}. 
    
    In this work, we seek to address this challenge by considering the online setting of the \ac{sac} problem, in which the \ac{inp} does not have information about future requests and must make an irrevocable decision (i.e., admit or reject) on a \ac{nsr} upon its reception.
    By recognizing the commonalities (and differences) between the considered \ac{osac} problem and the \ac{omdkp}, we develop an online algorithm for the considered problem. 
    
    The proposed algorithm admits \acp{nsr} onto the network infrastructure based on a reservation-based admission policy to maximize the long-term revenue received from admitted requests, while also offering a lower average resource utilization to potentially increase the profits of \acp{inp}. 
    The uncertainties in arrival rates, resource requirements, and the expected revenue of \acp{nsr} are addressed in an online manner in the algorithm.
    Due to a lack of approaches that jointly consider the online and multi-dimensional scenario for the \ac{sac} problem, to evaluate the performance of our approach, we compare it to an online greedy policy, which admits \acp{nsr} without consideration of the revenue associated with each request.
    Our contributions can be summarized as follows:
    
    \begin{itemize}
        \item \textcolor{black}{We model the \ac{osac} problem as an \acs{omdkp} with unknown resource demands and lifetimes and where the objective is to maximize the long-term revenue received from \acp{nsr}, while respecting the capacity constraints of the system resources.} 
        \item To address the \ac{osac} problem, we use two approaches, \ac{linrp} and \ac{exprp}, and propose an online algorithm that utilizes either of the two policies to perform admission control of \acp{nsr} in the considered scenario.
        \item Through extensive simulations, we evaluate the performance of the online algorithms in terms of the average resource utilization, the \ac{acr}, and total revenue gained compared to an online greedy solution.
    \end{itemize}
    
    The remainder of this paper is organized as follows. 
    In the next section, we present related works. 
    Section~\ref{sec:systemodel} presents the system model, describes the \ac{sac} problem, and formulates it as an \acs{omdkp}.
    The online algorithm and the reservation-based policies used to solve the \ac{omdkp} formulation are described in Section \ref{sec:onlineSAC}.
    In Section \ref{sec:perfeval}, we evaluate our proposed approach and compare it to an online greedy admission policy, \ac{fcfs}.
    We conclude the paper in Section~\ref{sec:conclusion}.
    
    \section{Related Work} \label{sec:relatedwork}
    The \ac{adc} problem in mobile network settings has been well-studied.
    However, previous works focused on devising mechanisms for the admission of mobile users to the network, as part of the \ac{rrm} process. 
    The emerging virtualization of the network infrastructure, and the softwarization of network functionality to support multi-tenancy through network slicing, increases the complexity of the \ac{adc}.
    As such, there is a need to explore new admission control mechanisms and approaches for such networks.
    In the literature, several works focus their efforts on applying \ac{ml} and \ac{rl} techniques to address the problem of \ac{adc}.
    To determine an optimal \ac{sac} policy, Bakri et al.~\cite{bakri2021} compare the performance of both online and offline solutions.
    Gholamipour et al.~\cite{OnlineAdmissionControl} formulate a joint online admission control and resource allocation problem based on an \ac{ilp}.
    Their solution considers the energy consumption of network nodes, as well as the workload uncertainties of sliced \acp{vnf} in their approach.
    In~\cite{articleONETS}, Sciancalepore et al. enable concurrent slice requests to be deployed over an \ac{inp}'s physical resources while maximizing multiplexing gains.
    Their approach focuses on addressing the \ac{sac} problem under demand uncertainty of network slices.
    Salvat et al.~\cite{10.1145/3281411.3281435} focus on the slice orchestration problem by jointly considering admission control and resource reservation.
    They propose solutions based on an optimal Benders decomposition method and a sub-optimal heuristic, to address the considered problem.
    Finally, Noroozi et al.~\cite{9013617} study the \ac{sac} problem, where each slice is made of \ac{ran} and \ac{cn} resources, and propose a sub-optimal two-step heuristic algorithm to maximize the total revenue gained from admitted slices.
    
    A key difference between the aforementioned works and the approach presented in this work is that we focus primarily on admission control and maximize the long-term revenue from the admitted slice requests.
    No previous work has considered the \ac{sac} problem under both the online and multi-dimensional settings.
    Following this, the presented approach can capture unbalanced demands across multiple resource dimensions to avoid over-utilizing a single resource by the accepted slices. 
    This is achieved by leveraging the concept of an \textit{admission cost} in the criteria of admission control decisions, where the cost to admit new \acp{nsr} to the infrastructure increases with the utilized resources and is an important factor in contextualizing the scarcity of any given resource in the network.
    
    
    \section{System Model and Problem Formulation} \label{sec:systemodel}
    
    \begin{table}
        \centering
        \caption{Summary of Notations}
        \begin{tabular}{lp{6.5cm}} 
            \toprule
            Symbol & Description \\
            \midrule
            $m$ & Number of resource types \\
            $\mathcal{H}$ & Index set of \acp{nsr} \\ 
            $\boldsymbol{r}_{h,j}$ & Requested amount of resource $j$ by \ac{nsr} $h$ \\
            $\delta_h$ & Lifetime of \ac{nsr} $h$\\
            $\tau_h$ & Timestamp of \ac{nsr} $h$\\
            $\pi_h$ & Revenue of \ac{nsr} $h$  \\
            $p_h$ & Unit value of \acp{nsr} $h$\\ 
            $\alpha_h$ & Coefficient vector of \acp{nsr} $h$\\ 
            $C_j$ & Aggregated capacity of resource $j$ \\
            $x_h$ & Decision variable on \ac{nsr} $h$\\ 
            $q^\text{lin}_{h,j}$, $q^\text{exp}_{h,j}$ & Normalized Resource Utilization of \ac{linrp} and \ac{exprp} policies for resource $j$ when \ac{nsr} $h$ enters the system \\ 
            $\phi^\text{lin}_h$, $\phi^\text{exp}_h$ & System Admission Cost of \ac{linrp} and \ac{exprp} policies when \ac{nsr} $h$ enters the system \\
            $\kappa_j$ & Resource heterogeneity ratio for resource $j$ \\
            $\theta$ & \acl{wtpr}\\
           \bottomrule
       \end{tabular}
        \label{tab:campus}
    \end{table}
    
    In this section, we formulate our problem in a typical network slicing scenario, in which an \ac{inp} owns the network resources and can lease them to \acp{st} for variable periods. 
    Thus, we describe the system model of the virtualized mobile network and formulate the \ac{sac} problem.
        
    \subsection{System Model}
    
    \begin{figure}
      \includegraphics[width=\columnwidth, height=8.5cm]{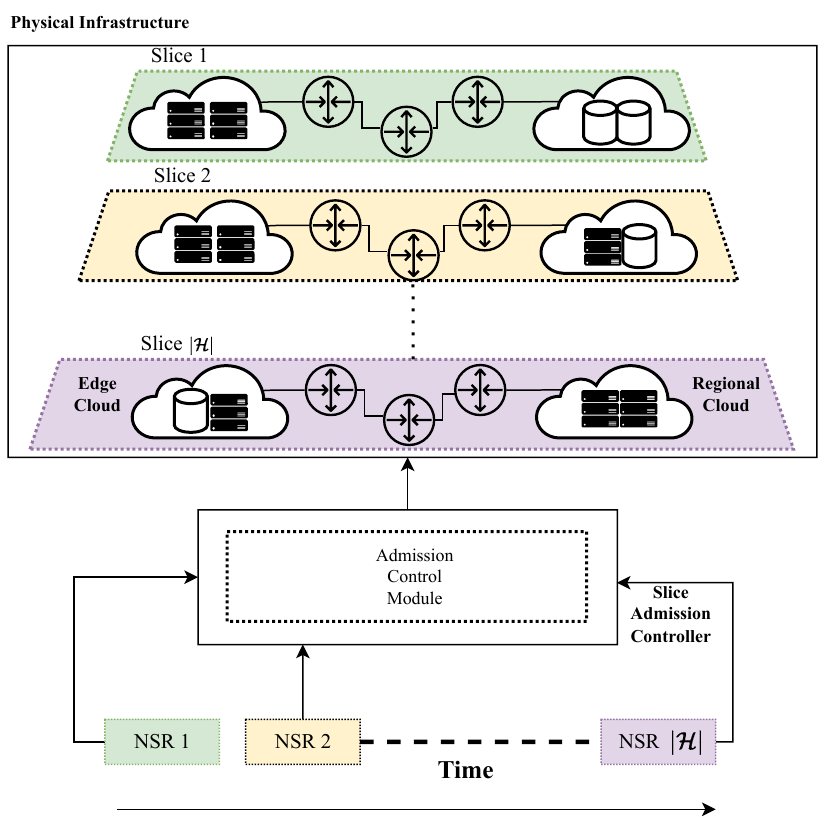}
      \caption{Overview of System}
      \label{fig:Architecture}  
    \end{figure}
    
    \textbf{Infrastructure Model}:
    We consider a scenario that consists of a sliced edge infrastructure made of \acp{bs} co-located with \acp{es}, which are equipped with an arbitrary amount of virtualized network and compute resources (e.g., bandwidth, \acs{cpu}, \acs{ram}, storage).
    In this scenario, an \ac{inp} owns and leases the physical network resources by dynamically allocating them to incoming \acp{nsr}.
    We assume that the physical resources of all edge servers in the system are pooled together to support heterogeneous \acp{ns} with different resource requirements.
    Therefore, in this work, the NSRs are admitted and executed on a single virtual infrastructure, physically distributed over a set of edge servers.
    In the considered scenario, \acp{st} seek to deploy and instantiate new slices (with their \acp{vnf}) on the infrastructure to provide their services, as shown in Figure~\ref{fig:Architecture}.
        
    In line with the time-varying network conditions at the network edge in \ac{nfv}-enabled mobile networks, we assume that time in the system is divided into consecutive intervals called \textit{time slots}.
    Each time slot is indexed with an element of the ordered index set $T = \{1, \ldots, |T|\}$.
    We assume that the sliced edge infrastructure offers $m$ different types of resources, whose amount does not change over time.
    We define $[m] = \{1,\ldots,m\} $ as an index set that identifies each of the $m$ resource types.
    For example, in this work, we assume that $m=3$, as the resource pool primarily consists of computing resources such as CPU, memory, and storage, which can be hosted by edge servers or a regional cloud provider.
    We define the variable $C_j\in\mathbb{R}_+$, with $j \in$ $[m]$, as the aggregate capacity of the $j$-th resource in the whole infrastructure.
    For example, in a scenario with $m=3$ types of resources, $C_1$, $C_2$, and $C_3$ can indicate the infrastructure-wide capacity of CPU, memory, and storage, respectively.
    
    \textbf{Slice Request Model}:
    Each of the consecutive tenant-generated \acp{nsr} is identified by the incremental index $h\in \mathcal{H}$, where $\mathcal{H} = \{1, \ldots, |\mathcal{H}|\}$ is defined as the ordered set of all \ac{nsr} indices.
    Furthermore, each \ac{nsr}'s features are summarized by a tuple $(\boldsymbol{r}_h, \delta_h, \tau_h, \pi_h)$, whose components are described hereafter.
    The variable $\boldsymbol{r}_h$ is an $m$-dimensional vector, where each component $r_{h,j}$ is the requested amount of a resource of type $j$ by an \ac{nsr} $h$.
    The variable $\delta_h \in \mathbb{N}$ represents the \textit{lifetime} of \acs{nsr} $h$, defined as the number of time slots the NSR needs to access the network resources.
    The variable $\tau_h \in \mathbb{N}$ represents the \textit{timestamp} of \acs{nsr} $h$, defined as the time slot index at which the request arrives to the scheduler.
    The variable $\pi_h \in \mathbb{R}_+$ represents the \textit{revenue} of \acs{nsr} $h$, defined as the economic benefit gained by the \ac{inp} from accepting the tenant's slice request and granting it its requested resources.
    We calculate it as $\pi_h = \delta_h p_h \boldsymbol{\alpha}_h \boldsymbol{r}_h $, with \textcolor{black}{$\lVert\boldsymbol{\alpha_{h,j}}\rVert_1 =1$}.
    The $j$-th component of the coefficient vector $\boldsymbol{\alpha} \in (0,1]^m$ represents the relative weight of the $j$-th resource to constitute a logical resource unit, and $p_h$ represents the \textit{unit value} of each logical resource in the \ac{nsr}.
    \textcolor{black}{The Manhattan-norm $\lVert\boldsymbol{\alpha_{h,j}}\rVert_1$ is used to represent the sum of coefficients of each resource $j$, for an \ac{nsr} $h$, and is given by $\sum\limits_{j \in [m]} \alpha_{h,j} = 1$.} 
    In this revenue model, we assume that a higher value of a coefficient represents a higher priority of the resource towards meeting the \ac{sla} of the \ac{nsr}.
    
    
    \subsection{Problem Formulation} \label{subsect:OMdKPForm}
    The considered dynamic \ac{sac} problem aims at maximizing the long-term revenue collected from admitting \acp{nsr} by optimizing the decisions on which \acp{nsr} should be admitted to the system while respecting the capacity requirements of the available resources. 
    
    Let us define $x_h$ as the \textit{decision variable} that takes a value of $1$, if the \ac{nsr} is admitted, and $0$ otherwise.
    We define $\boldsymbol{x}=(x_1,\ldots,x_{|\mathcal{H}|})$ as the \textit{decision vector}, where $\boldsymbol{x}\in\{0,1\}^{|\mathcal{H}|}$.
    Let us define the \textit{indicator function} $\mathbbm{1}_A:T\to\{0,1\}$ so that $\mathbbm{1}_A(t) = 1 \iff t\in A$ and $\mathbbm{1}_A(t) = 0 \iff t\notin A$. 
    We can now define the \ac{sac} problem as follows.
    \begin{maxi!}|s|[2]                   
        {\boldsymbol{x} \in \{0,1\}^{|\mathcal{H}|}}
        {\sum\limits_{h \in \mathcal{H}} \pi_{h}{x_{h}}
        \label{eq:eq1}}
        {\label{eq:Example1}}             
        {}                                
        \addConstraint \sum\limits_{h \in \mathcal{H}} r_{h,j} x_h \mathbbm{1}_{[\tau_h, \tau_h+\delta_h]}(t) \leq C_j,  \label{eq:con1}
        \addConstraint \qquad\qquad\qquad\qquad \forall j \in [m], \forall t \in T \nonumber \label{eq:con1.1}
    \end{maxi!}
    
    In this problem, constraint~\ref{eq:con1} ensures that at any time slot and for any resource, the system allows only a number of NSRs whose aggregate resources are less than the aggregate system capacity.
    \textcolor{black}{We observe that the stated \ac{sac} problem is NP-hard \cite{8057230}, as it can be reduced to a variant of the \ac{mdkp} problem, which is also NP-hard.}
    The literature offers several methods to approximate the solution of an NP-hard problem through algorithms that have polynomial complexity.
    However, the main limitation of this \textit{offline} formulation of the \ac{sac} problem is that, to compute the optimal solution, the scheduler must know the whole sequence of arriving \acp{nsr}, including those who will arrive in the future, which is unfeasible in real-world scenarios.
    \textcolor{black}{The lack of future information when deciding on the admission of \acp{nsr} makes the considered \ac{sac} intractable \cite{8678397}. }
    Hence, to enable real-world systems to approximate the solution of the stated offline \ac{sac} problem, we design a scheduler that uses an \textit{online} algorithm for \acl{sac} to optimize revenues.

    
    \section{Online Slice Admission Control} \label{sec:onlineSAC}
    
    We consider the online version of the \ac{sac} problem, where the unknown inputs are the multi-dimensional resource requirements (or demand) of the \acp{nsr}, and their corresponding revenue. 
    In such a problem, the goal is to design an online algorithm that determines how sequentially arriving requests are accepted, based only on the currently available information.
    \textcolor{black}{Specifically, we consider the scenario in which arrving \acp{nsr} cannot be buffered in the scheduler \textit{queue}, and therefore a decision must be made on it's admission immediately and before the end of the current time-slot, as this would enable immediate resource allocation for the arriving \ac{nsr}}. 
    Finding a policy that can solve the \ac{sac} problem in polynomial time while considering the unpredictable nature in which requests can arrive, is challenging. 
    
    We address this challenge based on the \ac{omdkp} formulation of the considered problem and leverage two policies\cite{Yang2021CompetitiveAF}, \ac{linrp} and \ac{exprp}, in our solution.
    The policies are based on online reservation functions that associate an implicit admission cost based on the utilization of each network resource.
    The intuition behind this is that by multiplying the current scarcity of each resource by the corresponding resource requirement of an \ac{nsr}, the cost of admitting a request is effectively evaluated in an online manner. 
    Furthermore, it is important to discourage high utilization of any resource in order to increase the chances of admitting higher valued future requests, as well as avoid performance degradation from resource sharing between co-located slices.
    
    \textcolor{black}{Let us define the \textit{\ac{wtpr}} $\theta$, as the ratio between the maximum and minimum price per resource unit per time slot the tenants are willing to pay to reserve resources on the provider's infrastructure.
    The main purpose of $\theta$ is to quantify the spread (i.e., the variation) in the \acp{nsr}' utilities, and, therefore, scenarios with a higher value of $\theta$ lead to higher revenues.
    However, such scenarios could likely have an impact on the admission rate due to the need for guaranteed \ac{qos} from high paying tenants.}
    In this work, we assume that the value of $\theta$ is fixed and known in advance (Section~\ref{sec:perfeval}).
    However, in real-world scenarios, the slice admission controller can learn the value of $\theta$ online by exploiting real-time information on sequentially arriving \acp{nsr} and by using data-driven algorithms.
    Let us define $u_{h,j}$ as the \textit{utilization} of the network resource $j$ after a \ac{nsr} $h$ has been accepted onto the infrastructure, and let us recall that $C_j$ is the total capacity of the $j$-th resource in the network.
    Finally, let us define the \textit{resource heterogeneity ratio} $\kappa_j = \frac{1}{C_j} \sum_{z\in[m]} C_z $, which captures the heterogeneity of the capacities of individual resources in the network (an important factor in the considered multi-dimensional setting).
    The details of the two policies are as follows.
    \begin{itemize}
    
        \item \textbf{\acl{linrp} (\ac{linrp})}:
            In this policy, the scarcity of each network resource increases linearly based on the normalized utilization of the resource.
            Hence, the higher the utilization of a given resource is, the higher is the admission cost of admitting a new \ac{nsr} that requests that resource.
            As a result, in periods where the overall network infrastructure has a higher load leading to higher resource utilization, the admission cost increases too to reserve the resources for higher valued requests.            
            Therefore, we define the \textit{normalized resource utilization} $q^\text{lin}_{h,j}$ and the \textit{system admission cost} $\phi^\text{lin}_h$ for \ac{linrp} in Equations~\ref{eq:linq} and~\ref{eq:linphi}, respectively.
            \begin{equation}
                \label{eq:linq}
                q^\text{lin}_{h,j} = \left\lfloor \frac{u_{h,j}}{C_j} \sqrt{\theta m} \right\rfloor
            \end{equation}
            \begin{equation}
                \label{eq:linphi}
                \phi^\text{lin}_h = \max_{{j} \in [m]} q_{h-1, j} \sqrt{\frac{2\kappa_j}{m}}r_{h, j}
            \end{equation}
            
        \item \textbf{\acl{exprp} (\ac{exprp})}: The second online reservation function considered is the \acl{exprp} (\ac{exprp}).
            Intuitively, the scarcity of each resource in the system increases following an exponential reservation function and is also based on normalized resource utilization.
            Therefore, we define the \textit{normalized resource utilization} $q^\text{exp}_{h,j}$ and the \textit{system admission cost} $\phi^\text{exp}_h$ for \ac{exprp} in Equations~\ref{eq:expq} and~\ref{eq:expphi}, respectively.
            \begin{equation}
                \label{eq:expq}
                q^\text{exp}_{h,j} = \left\lfloor \frac{u_{h,j}}{C_j} \log{(\theta \kappa_j)} \right\rfloor
            \end{equation}
            \begin{equation}
                \label{eq:expphi}
                \phi^\text{exp}_h = \sum\limits_{j=1}^m (2^{q_{h-1,j}} - 1)r_{h, j}
            \end{equation}
            Due to the exponential increase in the scarcity of resources, the \ac{exprp} approach can be considered as a more conservative approach, compared to \ac{linrp}, which could be suitable for situations with very high resource utilization to ensure only the highest valued requests are accepted. 
    \end{itemize}
    
    \begin{algorithm}
        \DontPrintSemicolon
        \SetAlgoLined
        \SetKwFor{Loop}{Loop}{}{EndLoop}
        \SetKwFunction{WaitNSR}{WaitNSR}
        \SetKwFunction{CheckFinishedNS}{CheckFinishedNS}
        \SetKwFunction{GetFinishedNS}{GetFinishedNS}
          \KwInput{$\theta$, $\kappa_j$, $C_j$, $\forall j\in[m]$;}
          \KwOutput{$x_h$;}
          $(h,u_0,q_0)\gets(0,0,0)$\;          
          \Loop{}{
          $h\gets h+1$\;
          $(\boldsymbol{r}_h, \delta_h, \pi_h) \gets$ \WaitNSR{} \;

          \tcp{\textcolor{black}{Update Admission Threshold (if LinRP)}}
          $\phi_h \gets \phi^\text{lin}_h = \max_{{j} \in [m]} q_{h-1, j} \sqrt{\frac{2\kappa_j}{m}}r_{h,j}$ \;

          \tcp{\textcolor{black}{Update Admission Threshold (if ExpRP)}}
          $\phi_h \gets \phi^\text{exp}_h = \sum\limits_{j=1}^m (2^{q_{h-1,j}} - 1)r_{h, j}$ \;

          $x_h \gets 0$ \;
          \If{$\pi_{h} \geq \phi_h$ and $r_{h,j} \leq C_j - u_{h-1,j}, \forall j\in[m]$}
            {
                $x_h \gets 1$ \;
            }
          \tcp{\textcolor{black}{Update Resource Utilization}}
          $u_{h,j} \gets u_{h-1,j} + x_h r_{h,j}, \forall j\in[m]$ \;
          \While{\CheckFinishedNS{}}{
            $\boldsymbol{r}_h \gets $ \GetFinishedNS{} \;
            \tcp{\textcolor{black}{Release Resources}}
            $u_{h,j} \gets u_{h,j} - r_{h,j}, \forall j\in[m]$ \;
            }
          \tcp{\textcolor{black}{Update Normalized Utilization (if LinRP)}}
          $q_{h,j} \gets q^\text{lin}_{h,j} = \left\lfloor \frac{u_{h,j}}{C_j} \sqrt{\theta m} \right\rfloor, \forall j\in[m]$  \;
          \tcp{\textcolor{black}{Update Normalized Utilization (if ExpRP)}}
          $q_{h,j} \gets q^\text{exp}_{h,j} = \left\lfloor \frac{\textit{u}_{h,j}}{C_j} \log{(\theta \kappa_j)} \right\rfloor, \forall j\in[m]$  \;
          \Return{$x_h$}\;
          }
        \caption{Online Slice Admission Control}
        \label{alg:OSACA}
    \end{algorithm}
    
    Our algorithm works in a time-slotted fashion in which each \acp{nsr} that arrives in a given time slot is handled in real-time, upon which they are either accepted and admitted onto the infrastructure or rejected immediately.
    We now describe the steps of Algorithm~\ref{alg:OSACA}.
    When a new \ac{nsr} arrives, we check its resource requirements (line 5). 
    Then, the admission threshold is updated based on the current scarcity of each resource with the requested resource.
    When the \ac{linrp} policy is applied, the scarcity is based on the resource with the higher scarcity or normalized utilization (line 6), while with the \ac{exprp} approach, the sum of scarcities in each dimension is summed up (line 7).
    A \ac{nsr} meets the criteria for admission when the value of the request is higher than the admission threshold for either policy and there is enough space in the system to accommodate the resource requirements of the request, otherwise the request is rejected (lines 8-10). 
    Based on the decision to accept or reject the \ac{nsr}, we update the resource utilization in the system (line 12).
    Since we consider that each admitted slice is in the system for a certain amount of time, at the end of each time slot we update the current resource utilization by releasing the allocated resources (line 15), and update the normalized utilization of the system too (lines 16-17).
    
    \section{Performance Evaluation} \label{sec:perfeval}
    \subsection{Simulation Setup} \label{subsec:SimSetup}
    
    We compare the performance of the implemented \ac{exprp} and \ac{linrp} policies with those of \ac{fcfs}, an online greedy policy that admits arriving \acp{nsr} onto the network infrastructure if there is sufficient capacity.
    We implemented the three policies and the simulated scenarios in Python and publicly released the source code on GitHub\footnote{\url{https://github.com/CDS-Bern/Online-Slice-Admission-Control}}. 
    As the literature does not offer public datasets on \acp{nsr}, we evaluate the performance of our proposed approach on a synthetic dataset $\mathcal{H}$ containing a sequence of $5\cdot10^7$ \ac{nsr} slots for each simulation.
    The size of this dataset is sufficient to estimate the average, long-term performance of the proposed approach with high confidence.
    We assume that the system contains $m=3$ resource types, namely CPU, RAM, and storage.
    Each \ac{nsr} in the dataset $h$ contains information about its duration $\delta_h$, requested resources $\boldsymbol{r}_h$, revenue $\pi_h$, and arrival timestamp $\tau_h$.
    We assume a time-slotted environment, where, during each time slot $t$, a random number $X_t$ of \acp{nsr} enters the system.
    We model $X_t$ as a Poisson process in which all $X_t, \forall t \in \{1,\ldots,|T|\}$ follow a Poisson distribution $\text{Pois}(\lambda)$ with identical arrival rate $\lambda=2$, where $\lambda$ is the average number of arrivals per slot.
    
    To evaluate how the compared policies perform under variable amounts of requested resources, we assume that each component of the resource vector $\boldsymbol{r}_h$ requested by a \ac{nsr} comes from a normalized uniform distribution $\mathcal{U}([0, 1])$.
    Furthermore, for the sake of simplicity, we assume that resource capacities are also normalized, so $C_{j}=1, \forall j \in [m]$.
    To simulate the impact of different durations on the system, we assume that the slice lifetimes $\delta_h$ are uniformly distributed as $\delta_h\sim\mathcal{U}(\{1, \zeta\})$, where $\zeta$ is the upper duration bound.
    To represent different economic conditions of tenants that issue \acp{nsr}, we model the slice \textit{unit value} $p_h$ as $p_h=1+(\sigma-1) Y$, where $Y$ is a random variable that follows a symmetric $\text{Beta}(\omega,\omega)$ distribution, and $\sigma\geq1$ is an \textit{economic scale} factor such that $p_h\in[1,\sigma]$.
    The parameter $\omega\in(0,+\infty)$ represents the \textit{economic inequality} of the tenants: when $\omega\to0$ an increasing share of tenants will request slices with unit values close to $1$ or to $\sigma$, while when $\omega\to+\infty$, tenants will offer increasingly similar unit values for their NSRs.
    We evaluated our approach in a set of scenarios, where $\omega\in\{0.05,0.1,\ldots,1\}$, $\sigma\in\{10,20\ldots,100\}$, and $\zeta\in\{10, 30, 100\}$.
    In these scenarios, the values of $\max_{i\in\mathcal{H}}{\delta_i p_i}=\sigma\zeta$ and $\min_{i\in\mathcal{H}}{\delta_i p_i}=1$.
    Therefore, the \ac{wtpr} $\theta$ that characterizes each scenario is $\theta=\sigma\zeta/1=\sigma\zeta$.
    Furthermore, considering that $C_j=1, \forall j \in [m]$, the resource heterogeneity ratio is $\forall j \in [m]: \kappa_j = m = 3$.
    
    The following three metrics are used to evaluate our approach:
    \begin{itemize}
        \item \textbf{Average Revenue Relative Gain}. We define the \textit{average revenue} as the ratio $\mu=\frac{1}{|\mathcal{H}|}\sum_{h\in\mathcal{H}}\pi_h$ between the total revenue and the number $|\mathcal{H}|$ of all received slice requests. We define the \textit{average revenue relative gain} for the two policies LinRP and ExpRP as $\frac{\mu_L-\mu_F}{\mu_F}$ and $\frac{\mu_E-\mu_F}{\mu_F}$, respectively, where $\mu_L$, $\mu_E$, and $\mu_F$ indicate the average revenues for LinRP, ExpRP, and FCFS, respectively. 
        \item \textbf{\acl{acr} Relative Gain}. We define the \textit{acceptance ratio} $\eta=n/|\mathcal{H}|$ as the ratio between the number $n$ of accepted slices and the number $|\mathcal{H}|$ of all received slice requests. 
        The \textit{acceptance ratio relative gain} for the two policies LinRP and ExpRP are defined as $\frac{\eta_L-\eta_F}{\eta_F}$ and $\frac{\eta_E-\eta_F}{\eta_F}$ respectively, where $\eta_L$, $\eta_E$, and $\eta_F$ indicate the acceptance ratios for the \ac{linrp}, \ac{exprp}, and \ac{fcfs} policies, respectively.
        \item \textbf{Average Resource Utilization Relative Gain}. We define the \textit{average resource utilization} $\rho=\frac{1}{|\mathcal{H}|}\sum_{h\in\mathcal{H}} \sum_{j\in[m]} u_{h,j}/C_j $ as the sum of the normalized utilization for all resources for all received slice requests divided by the number $|\mathcal{H}|$ of all received slice requests.
        The \textit{average resource utilization relative gain} for the two policies LinRP and ExpRP are defined as $\frac{\rho_L-\rho_F}{\rho_F}$ and $\frac{\rho_E-\rho_F}{\rho_F}$ respectively, where $\rho_L$, $\rho_E$, and $\rho_F$ indicate the average resource utilization for the \ac{linrp}, \ac{exprp}, and \ac{fcfs} policies, respectively.

    \end{itemize}
    
    \subsection{Simulation Results} \label{subsec:Results}
            \begin{figure*}
        \centering
            \begin{subfigure}[t]{0.31\textwidth}
                \centering
    \begin{tikzpicture} 
    \tikzstyle{every node}=[font=\scriptsize]
    \pgfplotsset{%
    compat=1.12,
    width=\textwidth,
    height=50mm
    }

    \begin{axis}
    [
        xlabel = {$\omega$},
        ylabel = {$\frac{\mu-\mu_F}{\mu_F}$},
        scaled y ticks=base 10:2,
        ymajorgrids=true,
        xmajorgrids=true,
        legend style={legend columns=6,at={(-0.15,1.12)},font=\scriptsize,anchor= south west}
    ]
        \addplot+
        plot [red, no markers, error bars/.cd, y dir=both, y explicit]
        table [x=unit_value_beta_params, y=linrp_rev_gain, col sep=comma, y error=y_error_lin] {img/rev10.csv};
        
        \addplot+
        plot [pink, no markers, error bars/.cd, y dir=both, y explicit]
        table [x=unit_value_beta_params, y=exprp_rev_gain, col sep=comma, y error=y_error_exp] {img/rev10.csv};
        
        \addplot+
        plot [blue, no markers, error bars/.cd, y dir=both, y explicit]
        table [x=unit_value_beta_params, y=linrp_rev_gain, col sep=comma, y error=y_error_lin] {img/rev30.csv};
        
        \addplot+
        plot [cyan, no markers, error bars/.cd, y dir=both, y explicit]
        table [x=unit_value_beta_params, y=exprp_rev_gain, col sep=comma, y error=y_error_exp] {img/rev30.csv};
        
        \addplot+
        plot [green, no markers, error bars/.cd, y dir=both, y explicit]
        table [x=unit_value_beta_params, y=linrp_rev_gain, col sep=comma, y error=y_error_lin] {img/rev100.csv};
        
        \addplot+
        plot [olive, no markers, error bars/.cd, y dir=both, y explicit]
        table [x=unit_value_beta_params, y=exprp_rev_gain, col sep=comma, y error=y_error_exp] {img/rev100.csv};
        
        \addlegendentry{$\zeta=10$ (LinRP)$\quad$}
        \addlegendentry{$\zeta=10$ (ExpRP)$\quad$}
        \addlegendentry{$\zeta=30$ (LinRP)$\quad$}
        \addlegendentry{$\zeta=30$ (ExpRP)$\quad$}
        \addlegendentry{$\zeta=100$ (LinRP)$\quad$}
        \addlegendentry{$\zeta=100$ (ExpRP)}
    \end{axis}
    \end{tikzpicture}
                \caption{Average Revenue Relative Gain}
                \label{fig:revenue_gain}
            \end{subfigure}~
            \begin{subfigure}[t]{0.31\textwidth}
                \centering
    \begin{tikzpicture}[trim axis left]
    \tikzstyle{every node}=[font=\scriptsize]
    \pgfplotsset{%
    width=\textwidth,
    compat=1.12,
    height=50mm
    }
 
    \begin{axis}
    [
        xlabel = {$\omega$},
        ylabel = {$\frac{\eta-\eta_F}{\eta_F}$},
        scaled y ticks=base 10:2,
        ymajorgrids=true,
        xmajorgrids=true,
        legend pos=south east,
    ]

        \addplot+
        plot [red, no markers, error bars/.cd, y dir=both, y explicit] 
        table [x = unit_value_beta_params, y = linrp_acr_gain, col sep = comma, y error = y_error_lin] {img/acr10.csv};
        
        \addplot+
        plot [pink, no markers, error bars/.cd, y dir=both, y explicit]
        table [x = unit_value_beta_params, y = exprp_acr_gain, col sep = comma, y error = y_error_exp] {img/acr10.csv};
        
        \addplot+
        plot [blue, no markers, error bars/.cd, y dir=both, y explicit] 
        table [x = unit_value_beta_params, y = linrp_acr_gain, col sep = comma, y error = y_error_lin] {img/acr30.csv};
        
        \addplot+
        plot [cyan, no markers, error bars/.cd, y dir=both, y explicit]
        table [x = unit_value_beta_params, y = exprp_acr_gain, col sep = comma, y error = y_error_exp] {img/acr30.csv};
        
        \addplot+
        plot [green, no markers, error bars/.cd, y dir=both, y explicit] 
        table [x = unit_value_beta_params, y = linrp_acr_gain, col sep = comma, y error = y_error_lin] {img/acr100.csv};
        
        \addplot+
        plot [olive, no markers, error bars/.cd, y dir=both, y explicit]
        table [x = unit_value_beta_params, y = exprp_acr_gain, col sep = comma, y error = y_error_exp] {img/acr100.csv};
        
     
    \end{axis}
 
    \end{tikzpicture}
    
                \caption{\acl{acr} Relative Gain}
                \label{fig:acceptance_ratio}
            \end{subfigure}~
            \begin{subfigure}[t]{0.31\textwidth}
                \centering
    \begin{tikzpicture}[trim axis left]
    \tikzstyle{every node}=[font=\scriptsize]
    \pgfplotsset{%
    compat=1.12,
    width=\textwidth,
    height=50mm
    }
 
    \begin{axis}
    [
        xlabel = {$\omega$},
        ylabel = {$\frac{\rho-\rho_F}{\rho_F}$},
        ymajorgrids=true,
        xmajorgrids=true,
        legend style={font=\scriptsize},
        legend pos=south east
    ]

        \addplot+
        plot [red, no markers, error bars/.cd, y dir=both, y explicit]
        table [x=unit_value_beta_params, y=linrp_util_gain, col sep=comma, y error=y_error_lin] {img/util10.csv};
        
        \addplot+
        plot [pink, no markers, error bars/.cd, y dir=both, y explicit]
        table [x=unit_value_beta_params, y=exprp_util_gain, col sep=comma, y error=y_error_exp] {img/util10.csv};
        
        \addplot+
        plot [blue, no markers, error bars/.cd, y dir=both, y explicit]
        table [x=unit_value_beta_params, y=linrp_util_gain, col sep=comma, y error=y_error_lin] {img/util30.csv};
        
        \addplot+
        plot [cyan, no markers, error bars/.cd, y dir=both, y explicit]
        table [x=unit_value_beta_params, y=exprp_util_gain, col sep=comma, y error=y_error_exp] {img/util30.csv};
        
        \addplot+
        plot [green, no markers, error bars/.cd, y dir=both, y explicit]
        table [x=unit_value_beta_params, y=linrp_util_gain, col sep=comma, y error=y_error_lin] {img/util100.csv};
        
        \addplot+
        plot [olive, no markers, error bars/.cd, y dir=both, y explicit]
        table [x=unit_value_beta_params, y=exprp_util_gain, col sep=comma, y error=y_error_exp] {img/util100.csv};
        
     
    \end{axis}
 
    \end{tikzpicture}
    
                \caption{Average Resource Utilization Relative Gain}
                \label{fig:mean_utilization}
            \end{subfigure}
        \caption{Relative Gain for \ac{linrp} and \ac{exprp} against \ac{fcfs} for different values of $\omega$ and $\sigma=10$. Confidence intervals at level 99.9\%.}
        \end{figure*}
    
    \subsubsection{Average Revenue Relative Gain}
    
    Figure~\ref{fig:revenue_gain} shows the relative gains of the \ac{linrp} (red, blue and green lines) and \ac{exprp} (light red, light blue, and light green) policies, compared to the greedy (\ac{fcfs}) policy for the mean revenue. 
    The results are based on the slice holding time $\delta_h$ of each incoming \acp{nsr}, which is bounded by a specific duration, i.e., $\zeta\in\{10, 30, 100\}$.
    From this, we can see that applying both policies in our algorithm leads to higher average revenue than the greedy policy across various upper bounds on the \ac{nsr} durations and for the range of unit values considered. 
    Specifically, it can be observed that in situations with higher economic inequality between tenants ($\omega \to 0$), the revenue gain is the highest (over 10\%) when using both policies.
    This is because, in such scenarios, the value of the accepted \acp{nsr} is higher due to the online reservation function that places a higher threshold on the values that can be accepted.

    \subsubsection{\acl{acr} Relative Gain} \label{subsubsect:acceptratio}

    The relative gains in the \ac{acr} by using the online reservation-based policies, are shown in Figure~\ref{fig:acceptance_ratio}. 
    We see that by utilizing the introduced policies in our approach, fewer \acp{nsr} are admitted compared to the greedy \ac{fcfs} approach for the considered range of upper-bounded \ac{nsr} durations times. 
    The performance of our solution improves when the average duration of requests increases and when the value of $\omega$ approaches 1.
    This is primarily due to the admission criteria of the reservation-based policies compared to that of the greedy policy, where the proposed solution rejects lower-valued requests at a higher rate compared to higher-valued slice requests with longer durations.
    This leads to a lower overall acceptance ratio for the considered parameters, but a higher revenue, as seen in the previous results for \textit{Average Revenue Relative Gain}.
    
    \subsubsection{Average Resource Relative Utilization} \label{subsubsect:utilization}
    
    It can be seen in Figure~\ref{fig:mean_utilization}, that across most of the evaluated upper-bounded duration and unit price parameters, the resource utilization is lower on average using our approach. 
    This is in line with the results from the acceptance ratio gain, which shows that based on the reservation-based policies our approach leads to a lower overall acceptance ratio, and hence, there are periods during the experiments when the resources are not fully utilized but new \acp{nsr} are rejected due to their revenue and the current utilization level of any given resource in the system.
    Achieving a lower average utilization is key to achieving the goal of maximizing the revenue for \acp{inp} as it ensures that the scarce resources are reserved for the \acp{nsr} that offer the most revenue.
    
    
    \section{Conclusion} \label{sec:conclusion}
    
    In this paper, we proposed an online approach that uses two novel reservation-based policies to solve the \ac{sac} problem in virtualized mobile networks. 
    We focus on the dynamic nature of such networks by considering an online and multidimensional scenario in which multi-resource \acp{nsr} arrive sequentially and must be admitted or rejected based on the provided information only, and in the absence of future information.
    \textcolor{black}{Through simulations, we show that our proposed method, which utilizes the \ac{linrp} or \ac{exprp} policy, outperforms a greedy \ac{fcfs} strategy in average revenue gain while accepting fewer \acp{nsr} and saving on the consumption of system resources.}
    
    \bibliographystyle{IEEEtran}
    \bibliography{IEEEabrv,references.bib}

    \end{document}